\documentclass[draft]{IEEEtran}
\IEEEoverridecommandlockouts
\usepackage{cite}
\usepackage{amsmath,amssymb,amsfonts}
\usepackage{algorithmic}
\usepackage{graphicx}
\usepackage{textcomp}
\usepackage[table,xcdraw]{xcolor}
\usepackage{xcolor}
\usepackage{multirow}
\usepackage{minted}
\begin{document}

\title{BoostNSift: A Query Boosting and Code Sifting Technique for Method Level Bug Localization}

\author{\IEEEauthorblockN{Abdul Razzaq}
\IEEEauthorblockA{\textit{University of Limerick}\\
Limerick, Ireland \\
abdul.razzaq@lero.ie}
\and
\IEEEauthorblockN{Jim Buckley}
\IEEEauthorblockA{
\textit{University of Limerick}\\
Limerick, Ireland \\
jim.buckley@ul.ie}
\and
\IEEEauthorblockN{James Vincent Patten}
\IEEEauthorblockA{
\textit{University of Limerick}\\
Limerick, Ireland \\
james.patten@ul.ie}
\and
\IEEEauthorblockN{Muslim Chochlov}
\IEEEauthorblockA{
\textit{University of Limerick}\\
Limerick, Ireland \\
muslim.chochlov@ul.ie}
\and
\IEEEauthorblockN{Ashish Rajendra Sai}
\IEEEauthorblockA{
\textit{University of Limerick}\\
Limerick, Ireland \\
ashish.sai@ul.ie}
}
\maketitle

\begin{abstract}
Locating bugs is an important, but effort-intensive and time-consuming task, when dealing with large-scale systems. To address this, Information Retrieval (IR) techniques are increasingly being used to suggest potential buggy source code locations, for given bug reports. While IR techniques are very scalable, in practice their effectiveness in accurately localizing bugs in a software system remains low. Results of empirical studies suggest that the effectiveness of bug localization techniques can be augmented by the configuration of queries used to locate buggy code. However, in most IR-based bug localization techniques, presented by researchers, the impact of the queries' configurations is not fully considered. In a similar vein, techniques consider all code elements as equally suspicious of being buggy while localizing bugs, but this is not always the case either.

In this paper, we present a new method-level, information-retrieval-based bug localization technique called ``BoostNSift''. BoostNSift exploits the important information in queries by `boost'ing that information, and then `sift's the identified code elements, based on a novel technique that emphasizes the code elements' specific relatedness to a bug report over its generic relatedness to all bug reports. To evaluate the performance of BoostNSift, we employed a state-of-the-art empirical design that has been commonly used for evaluating file level IR-based bug localization techniques: 6851 bugs are selected from commonly used Eclipse, AspectJ, SWT, and ZXing benchmarks and made openly available for method-level analyses. The performance of BoostNSift is compared with the openly-available state-of-the-art IR-based BugLocator, BLUiR, and BLIA techniques. Experiments show that BoostNSift improves on BLUiR by up to 324\%, on BugLocator by up to 297\%, and on BLIA up to 120\%, in terms of Mean Reciprocal Rank (MRR). Similar improvements are observed in terms of Mean Average Precision (MAP) and Top-N evaluation measures.

\end{abstract}


\section{Introduction}
Debugging code, in response to bug-reports from users, while challenging and time consuming, is an essential activity that developers must engage in~\cite{anvik2005coping}. Before any reported bug can be fixed, developers must first locate the buggy code elements in the software system(s), an activity that can often lead to significant delays in the bug fixing process~\cite{anvik2005coping,jeong2009improving,parnin2011automated}. It is therefore unsurprising to see a demand for better automated bug localization solutions \cite{kochhar2016practitioners} and many bug localization techniques have been proposed by researchers to help developers in locating buggy code elements \cite{youm2017improved,zhou2012should,kim2013should,ye2016mapping,saha2013improving}. These techniques operate using a variety of approaches: textual or structural analysis of the source code \cite{wen2016locus,zhang2016baha,zhang2019finelocator}, historic analysis of software change information \cite{youm2017improved,wang2014version} and dynamic analysis of pass/fail test-case information \cite{le2015information}. As bug reports are usually formulated in natural language and the source code also includes large amounts of comments and identifiers, Information-Retrieval (IR) based bug localization techniques are frequently proposed for bug localization \cite{youm2017improved,zhou2012should,kim2013should,ye2016mapping,saha2013improving, zhang2019finelocator,wen2016locus,dit2015configuring,biggers2014configuring,shi2014empirical,moreno2015query,dit2015supporting,tantithamthavorn2018impact,chaparro2017using}.

IR-based bug localization techniques exploit the knowledge encoded in bug reports and the associated source code to link bug reports with corresponding buggy code elements. This means the performance of IR-based techniques are largely dependent upon the queries used as input to the bug localization process \cite{mills2017predicting,thomas2013impact} and the possibly-buggy code elements \cite{catal2009systematic}. For this reason associated branches of research are dedicated to identifying better configurations for queries \cite{mills2017predicting,thomas2013impact,biggers2014configuring} and on assessing the code elements \cite{catal2009systematic,rahman2016improved,takahashi2018preliminary}. The former studies focus on the impact that each constituent component of a query has on the performance of techniques \cite{moreno2015query,mills2017predicting} or on pre-processing of queries to improve their quality for bug localization \cite{mills2017predicting}. The latter studies focus on scoping the ranked subset returned by bug-localization approaches to a certain number of elements, but they do not endeavour to \textit{sift} the ranked-list \cite{le2015information,poshyvanyk2007feature,dit2013integrating}. While the findings of these studies confirm the impact that query configurations and code element filtration can have on the performance of bug localization techniques \cite{dit2015configuring,biggers2014configuring,tantithamthavorn2018impact,moreno2015query,thomas2013impact}, existing bug localization techniques do not fully exploit these in terms of weighting important queries components and \textit{sifting} the code elements in ranked-list to improve the ranking.

Therefore this paper proposes a novel method-level IR-based bug localization technique \textit{BoostNSift} that incorporates query optimization and result filtering in a manner designed to improve the rankings. BoostNSift is comprised of three components: QueryBooster, which leverages the structure of bug reports to boost elements of the query for bug localization; BM25, a more user-intuitive, non-binary IR-model which is less susceptible to document sizes, to estimate the relevance of code elements to a given (boosted) query, and CodeSifter which performs sifting analysis on the source code elements in order to refine the generated output. To perform the sifting analysis, CodeSifter compares the score of a code element returned from analysis of a specific bug report to the score returned from analysis with respect to a bug report aggregating all bug reports: The hypothesis being, the greater the difference in score, the more specifically relevant the returned code. BoostNSift therefore offers a lightweight analysis for bug localization that requires only bug reports and source code as inputs, and boosts queries/sifts resultant code for locating bugs. 

Additionally, as most of the existing bug localization techniques locate buggy code at file level in the source code \cite{youm2017improved,zhou2012should,kim2013should,ye2016mapping,saha2013improving} developers may be required to review a lot of code in order to find the lines that contain the bug \cite{le2015information}. If the associated code was instead located at method level, as suggested by Kochhar et al.\cite{kochhar2016practitioners}, much less effort would be required from developers. We therefore adopt a method-level granularity \cite{youm2017improved,zhang2019finelocator,tantithamthavorn2018impact,rahman2016improved}.

To evaluate the effectiveness of BoostNSift, we compared it with existing commonly-used, openly-available, state-of-the-art IR-based bug localization techniques, employing commonly-used benchmarks, and evaluation measures. As the focus here is on method-level analysis \cite{zhang2019finelocator}, we derived a method-level dataset from existing empirical designs. Specifically, we compared BoostNSift with the openly-available BugLocator \cite{zhou2012should}, BLUiR \cite{saha2013improving} and BLIA \cite{youm2017improved} techniques employing 6851 bugs, at method level, selected from commonly-used Eclipse, AspectJ, SWT, and ZXing Benchmarks. The implementation of BoostNSift and method-level dataset produced in this paper can be accessed by following the link: https://github.com/razi-rechan/boostNsift.

The remainder of this paper is as follows: Section~\ref{sec2:work} presents a brief literature review that demonstrates the impact of queries and output refinement on the performance of bug localization techniques. Section~\ref{sec3:approach} presents the proposed approach, detailing the steps performed by each component of BoostNSift and illustrates those steps by an example. Then, Section~\ref{sec4:design} presents the research questions and details the empirical design used to answer each question: The associated results are presented in Section~\ref{sec5:results}. Later, Section~\ref{sec6:discussion} discusses the results, describing the efficacy of BoostNSift. Threats to validity are discussed in Section~\ref{sec7:threats}. Finally, conclusions and future work are presented in Section~\ref{sec8:conc}.

\section{Background}\label{sec2:work}

\subsection{Bug Localization Process}

When a developer or user finds a bug in the software, (s)he usually describes it in a bug report and submits that to a bug tracking repository. A bug manager triages the bug and assigns it to a developer who has to fix the bug. After accepting the bug, the developer usually needs to locate the buggy source code elements before fixing them. This is known as bug localization \cite{ali2013trustrace} and is the core focus of this paper. A bug localization technique aims to locate the source code elements in a software system that correctly correspond to a bug report \cite{borg2014recovering}.

\subsection{Impact of Input-and-Output Refinement}

The representations of inputs (typically bug reports, and code) affect the efficacy of IR-based bug localization techniques, as suggested by \cite{thomas2013impact}. Common to all IR-based techniques, the following input considerations have to be decided:
\begin{enumerate}
    \item Which parts of the bug report\footnote{https://bugs.eclipse.org/bugs/show\_bug.cgi?id=462629} should be considered: the title, the description, comments, or all?
    \item Which parts of the source code should be considered: the comments, (various) identifier names, and-or literals?
    \item How should the source code and bug report text be pre-processed? For example, should words to be stemmed into their root form?
\end{enumerate}

Several studies have assessed the impact of these different configurations on the performance of bug localization techniques \cite{dit2015configuring,biggers2014configuring,tantithamthavorn2018impact,moreno2015query,thomas2013impact} and Table \ref{tab1:Configurations} presents selected results from those studies. The \textit{Worst} and \textit{Best} columns present the worst and best performance of the technique using different configuration settings. The (large) impact of these input configurations is evident from the table. 
\begin{table*}[]
\centering
\caption{Impact of Input Configurations}
\label{tab1:Configurations}
\setlength{\tabcolsep}{2pt}
\renewcommand{\arraystretch}{1.3}
\begin{tabular}{|cccc|}
\hline
\multicolumn{1}{|c}{\cellcolor[HTML]{C0C0C0}\textbf{Study}} &
  \cellcolor[HTML]{C0C0C0}\textbf{\# Configurations} &
  \cellcolor[HTML]{C0C0C0}\textbf{Worst} &
  \cellcolor[HTML]{C0C0C0}\textbf{Best} \\ \hline \hline
\multicolumn{1}{|l|}{\cellcolor[HTML]{EFEFEF} Thomas et al. [29]}           & 3172  & 1.1\% of top-20 suggestions were relevant code     & 55\% of top-20 suggestions were relevant code    \\ \hline
\multicolumn{1}{|l|}{\cellcolor[HTML]{EFEFEF} Tantithamthavorn et al. [28]} & 3172  & 0.3\% of top-20 suggestions were relevant code     & 38\% of top-20 suggestions were relevant code    \\ \hline
\multicolumn{1}{|l|}{\cellcolor[HTML]{EFEFEF} Biggers et al. [3]}          & $>$1000 & Highest-ranked relevant code element: 11,348th & Highest-ranked relevant code element: 28th   \\ \hline
\multicolumn{1}{|l|}{\cellcolor[HTML]{EFEFEF} Shi et al. [26]}              & 6300  & Highest-ranked relevant code element: 6369th   & Highest-ranked relevant code element: 4376th \\ \hline
\multicolumn{1}{|l|}{\cellcolor[HTML]{EFEFEF} Moreno et al. [16]}           & 21    & 0\% queries retrieved some relevant code       & 43\% queries retrieved some relevant code   \\ \hline
\end{tabular}
\end{table*}

\begin{table}[h]
\caption{How this work compares to existing similar research}
\label{tbl:comparison_existing}
\setlength{\tabcolsep}{2pt}
\begin{tabular}{l|cccc}
\hline
                         & \begin{tabular}[c]{@{}c@{}}Input/output \\ components \\ configurations\end{tabular} & \begin{tabular}[c]{@{}c@{}}IR parameters' \\ configurations\end{tabular} & \begin{tabular}[c]{@{}c@{}}Components' \\ weights\end{tabular} & \begin{tabular}[c]{@{}c@{}}Sifting \\ the output\end{tabular} \\ \hline
Thomas et al.            & +                                                                                    & +                                                                        &                                                                &                                                               \\
Tantitthamthavorn et al. &                                                                                      & +                                                                        &                                                                &                                                               \\
Biggers et al.           & +                                                                                    & +                                                                        &                                                                &                                                               \\
Shi et al.               &                                                                                      & +                                                                        &                                                                &                                                               \\
Moreno et al.            & +                                                                                    & +                                                                        &                                                                &                                                               \\ \hline
Our approach             & +                                                                                    & +                                                                        & +                                                              & +                                                             \\ \hline
\end{tabular}
\end{table}

Specifically, for the bug reports themselves, Tantithamthavorn et al. \cite{tantithamthavorn2018impact}, Thomas et al. \cite{thomas2013impact} and Biggers et al. \cite{biggers2014configuring} have all highlighted the different performance achievable when title and description are used. Over all the studies, the commonly-identified best input configuration is to use the title and descriptions in the query, and to use comments, (all) identifier-names, and literals to represent code elements. For pre-processing they employed stop-word removal, identifier splitting, term normalization and stemming steps \cite{dit2015configuring,biggers2014configuring,tantithamthavorn2018impact,moreno2015query,thomas2013impact}. However, none of these studies has considered adjusting the impact of each of the parameters: for example, using title and description together, but with different weights.

Similar to inputs, output processing also seems to impact the performance of bug localization techniques and several techniques filter specific code elements from the outcome returned by techniques \cite{youm2017improved,dit2013integrating} to improve their overall results. For example, Dit et al.\cite{dit2013integrating} suggested that filtering the top/bottom K number of elements in this fashion (where K is some artificial threshold), from the result-list of an IR-based technique can improve its results up to 33\%. 

So it seems that embedding input/output processing in bug localization improves the performance of techniques. While existing research assesses the impact of some input/output components, individually and in combination, it does not fully leverage them (unlike our work) to improve bug localization techniques, as illustrated in Table~\ref{tbl:comparison_existing}. This paper addresses and evaluates that concern.

\section{Proposed Approach}\label{sec3:approach}

\begin{figure}[htbp]
\centerline{\includegraphics[width=0.45\textwidth]{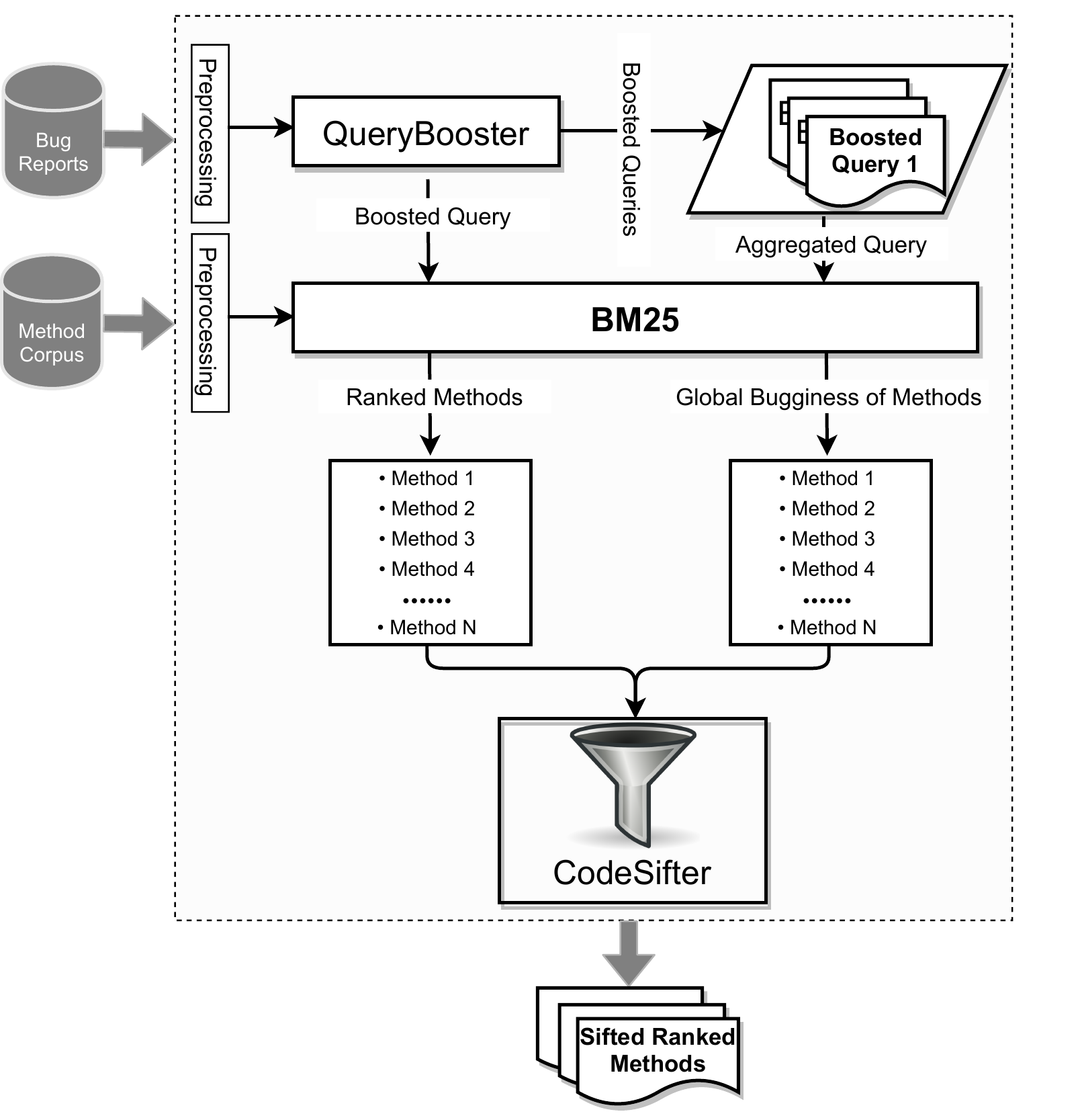}}
\caption{Architecture of BoostNSift}
\label{fig2:Architecture}
\end{figure}

The overall architecture of BoostNSift is presented in Figure \ref{fig2:Architecture}. It comprises of three core components with one pre-processing step. BoostNSift's inputs include the bug reports and the source code belonging to the software system where bugs are to be localized. We select the best identified constituents of bug reports and code elements to represent queries and the method-level, source code corpus, respectively. As a recent study by Youm et al. \cite{youm2017improved} found a positive impact of comments on the performance for bug localization, we include comments as well as bug descriptions and titles in our query input. These inputs are then pre-processed, following the commonly best-identified configurations, over several studies \cite{dit2015configuring,biggers2014configuring,tantithamthavorn2018impact,moreno2015query,thomas2013impact}:
\begin{enumerate}
    \item Identifier Splitting: Splitting code identifiers into their constituent words, and then keeping code identifiers in their original form as well as their split form. For example, the constituent words of `addWidget' are `add' and `Widget', with all three retained in the query;
    \item Term Normalization: Normalizing terms by converting them to a uniform case (lower in our case);
    \item Term Filtration: Filtering terms by removing stop-words (e.g. `is', `the') and-or common programming language keywords (e.g. `if', `else');
    \item Stemming Terms: To reduce words to their inflectional roots (e.g. `protection', `protective', `protected' are converted to `protect').
\end{enumerate}

For the sake of uniformity, the above four pre-processing steps are applied to both query and code inputs.

\subsection{QueryBooster}\label{sec3.1:qb}

Once the inputs are pre-processed, QueryBooster `boosts' each bug report. In this step, the structure of the textual parts of a bug report is exploited and the title, description, and comments of the bug reports are all incorporated. We contend, based on the findings of several studies \cite{dit2015configuring,biggers2014configuring,tantithamthavorn2018impact,moreno2015query,thomas2013impact} that each of these parts does not have the same impact.

Hence we performed a preliminary assessment of different weighting schemes on AspectJ (see Table \ref{tab3:Setting}). 
AspectJ was selected as per Saha et al.\cite{saha2013improving} and because it was a large system in our gold-set. For each constituent of the query, we tested a range of weights from 0.5 to 4.0 in 0.1 step-sizes: The MRR and MAP results of query boosting with the best three and worst three query configurations are shown in Table \ref{tab3:Setting}. 

\begin{table}[]
\centering
\caption{The best three and worst three configurations of BoostNSift in Query Boosting}
\label{tab3:Setting}
\setlength{\tabcolsep}{2pt}
\renewcommand{\arraystretch}{1.3}
\begin{tabular}{|l|lllll|lllll|}
\hline
\rowcolor[HTML]{C0C0C0} 
\multicolumn{1}{|c|}{\cellcolor[HTML]{C0C0C0}\textbf{Systems}} &
  \multicolumn{5}{c|}{\cellcolor[HTML]{C0C0C0}\textbf{Best}} &
  \multicolumn{5}{c|}{\cellcolor[HTML]{C0C0C0}\textbf{Worst}} \\ \hline \hline
\rowcolor[HTML]{EFEFEF} 
                          &$\alpha$*  & $\beta$* & $\gamma$*  & MRR   & MAP   & $\alpha$*  & $\beta$*  &$\gamma$*  & MRR   & \multicolumn{1}{l|}{\cellcolor[HTML]{EFEFEF}MAP} \\ \hline
                          & 3.0   & 1.0  & 2.0   & 0.14  & 0.091 & 0.5 & 3.3 & 3.4 & 0.082 & \multicolumn{1}{l|}{0.047}                       \\ \cline{2-11} 
                          & 2.4 & 1.0  & 1.2 & 0.14  & 0.091 & 0.7 & 3.3 & 3.4 & 0.104 & \multicolumn{1}{l|}{0.091}                       \\ \cline{2-11} 
\multirow{-3}{*}{AspectJ} & 3.1 & 1.0  & 1.0   & 0.139 & 0.09  & 0.6 & 3.3 & 3.3 & 0.126 & \multicolumn{1}{l|}{0.078}                       \\ \hline
\end{tabular}
\\
*$\alpha$=bug title's weight, $\beta$=bug descriptions' weight, $\gamma$=bug comments' weight
\end{table}

As guided by this assessment, BoostNSift weights the title, comments and descriptions of the bug reports with ratio 3:2:1, respectively. Boosted bug reports are also aggregated into a `grand' query, to be used in source code sifting (in Section~\ref{sec3:code-sifter}).
\subsection{BM25}
Boosted queries and the aggregated boosted query are then used as an input to an information retrieval model. The second input is the pre-processed method-corpus build in the pre-processing step. This research employs BM25 (Best Match 25) as an information retrieval model \cite{robertson2009probabilistic}; recently adopted as the default IR model in Lucene\footnote{https://lucene.apache.org/}.   We employed BM25 because it better corresponds to user intuitions of relevance than traditional TF*IDF (Term Frequency - Inverse Document Frequency) approaches \cite{saha2013improving}. For example, if a method mentions `widget' twelve times it is probably not twice as relevant as a method mentioning `widget' six times. Similarly, a term occurring in 100 methods is not twice as discriminating as a term occurring in 200 methods. To control these ranges of relevance, BM25 defines a saturation point that TF values cannot exceed. TF of BM25 is also sensitive to the method length. For example, large methods have a higher probability of getting query hits than small methods. But a bug can also occur in a small method. Hence, TF in BM25 is defined as follows:
\begin{equation}
TFi = ((k + 1) * ti) / (k * (1.0 - b + b * |m|/avg|M|) + ti)
\end{equation}

Where $ti$ represents the frequency (count) of a term in method $m$. $|m|/avg|M|$ in equation (1) is the current method length divided by the average method length. \textit{k} and \textit{b} are the tuning parameters used to define the saturation points for term frequency and method length, respectively. This means that shorter methods hit the saturation point faster on the best possible $TF$ score. In this paper, for most of the systems, we keep the default settings of Lucene for BM25 where \textit{k}=1.2 and \textit{b}=0.75 (see Section~\ref{sec4:eval-measures} for the exception). 

\subsection{CodeSifter} \label{sec3:code-sifter}

After potential code has been identified via BM25, the CodeSifter component of BoostNSift isolates the likely buggy elements specific to a bug from the ranked-lists generated by BM25. To perform a similar analysis, existing techniques use filtration strategies to isolate the code lying towards the top of the ranked-list. Those strategies are either based on an artificial threshold value of similarity score assigned by the IR-model to each code element \cite{ali2013trustrace} or a constant number K used to select the top K elements in the ranked-list \cite{youm2017improved}. The former practice, however, does not guarantee a size-free assessment of bug localization techniques because such a threshold will keep changing from system to system and technique to technique. Likewise, the latter technique of selecting the top K elements will be highly sensitive to the total elements in the system and the number of elements retrieved against a bug report \cite{razzaq2019empirical}. Another strategy, mostly adopted by techniques that perform hybrid analyses, is to combine multiple types of analyses to filter the results of one type of analysis with others [27, 28]. The results of these techniques emphasize the importance of such filtrations [5, 27, 28]. However, they require additional sources of information and processing. For example, in Le et al.\cite{le2015information}, to filter IR-based analysis results, spectrum based traces information and processing are required.

To filter lesser-related code elements, BoostNSift adopts a novel sifting method where it measures the bugginess of each code element to a specific bug report by comparing its global and local score. Its global score is the score of that element against an all-encompassing, boosted query (the aggregation of all boosted queries). Its local score is the score of the element to an individual, boosted bug report. A source code method is said to be buggy for a specific bug report if its score for that bug report is higher than its score on the aggregated boosted query. After sifting, we also filter the results for code elements for zero local scores and then rank the lists in descending-match order. 


\subsection{Illustrative Example}
\begin{listing}
\begin{minted}[breaklines,fontsize=\small]{java}
Deck() {
   c = new ArrayList<Card>();

   for (short a=0; a<=3; a++) {
    	for (short b=0; b<=12; b++) {
     	   //add cards to deck
     	   c.add(new Card(a,b)); 
     	}
   }
   shuffle();
}

Card(short s, short r) {
   //rank of card created: Ace, two, King
   this.r=r; 
   //suit: Spade, Club, Heart, Diamond
   this.s=s; 
}

Hand(Deck d) {
   v = new int[6];
   c = new Card[5];
   for (int x=0; x<5; x++) {
       c[x] = d.drawFromDeck();
   }
   HighestCardOnlyHand=true;
   getRanksOfCards();
   getPairsTripletsFours();
   Flush=isFlush();
   Straight=isStraight();
   setTypeOfHand();
   setHighestRankedCardValues();
}
\end{minted}
\caption{Toy Example with three methods}
\label{listing:toy}
\end{listing}

\begin{table}[h]
\centering
\caption{Bug Descriptions}
\label{tab:bugs}
\begin{tabular}{|l|l|p{6cm}|}
 \hline
1 & Title & Shuffling issue \\
 & Description & Deck is incorrectly shuffled: cards still in sequential order \\ \hline
2 & Title & Triplet-Flush Issue \\
 & Description & I had a flush but I seem to have lost to a triplet? \\ \hline
3 & Title & Ordering of Pairs \\
 & Description & We both had a pair and mine was higher, but my opponent won \\ \hline
\end{tabular}

\end{table}

\begin{table}[]
\centering
\caption{Query formation}
\label{tab:queryFormation}
\begin{tabular}{|l|l|p{6cm}|}
\hline
1 & Title & Shuffle \\
 & Description & shuffle, deck, incorrect, wrong, card, sequence, order \\ \hline
2 & Title & Triplet, three, flush \\
 & Description & flush, triplet, three, lost, defeat \\ \hline
3 & Title & order, rank, pair, two \\
 & Description & order, rank, pair, two,   high, opponent, won, victory \\ \hline
\end{tabular}

\end{table}

In this subsection, we provide an illustrative toy example to better explain BoostNShift. The source code for the illustration contains three methods: Deck, Card and Hand, as manifested in Listing \ref{listing:toy}. The bug reports for this example are present in Table \ref{tab:bugs}. In this case, the title and title description from the bug report are used to construct the query, and Table \ref{tab:queryFormation} contains the query for each of the bug reports, after synonym inclusion/stemming. Finally, the encompassing query over all titles and descriptions would be the following \{deck, incorrect, wrong, shuffle, card, sequence, order, flush, lost, defeat, triplet, three, pair, two, high, rank, opponent, won, victory\}.

BoostNSift weights bug titles three times higher than bug descriptions. In this toy example, it is apparent that the titles of the bugs are much more specific in terms of bug localization: ‘shuffle’ is mentioned only in the ‘Deck’s constructor, and the terms in the 2nd and 3rd bug titles are nearly exclusively mentioned in the constructor for ‘Hand’: all appropriate places to begin a code search for the respective bugs. So, in this case, it appears appropriate that bug titles should be given more weight than (or ‘boost’ed over) bug descriptions.

To illustrate the ‘Sift’ component, let’s consider the whole-bug "descriptors" (titles and descriptions) and calculate the approximation: "word-overlaps-between-descriptor-and-code/line-of-code" to report on the similarity between queries and methods. See Table \ref{tab:resultsDescriptors} for the results for each of the descriptors, and the results for the cumulative/global descriptor.

\begin{table}[]
\centering
\caption{Results for the descriptors}
\begin{tabular}{|l|l|l|l|} \hline
Descriptor & Deck & Card & Hand  \\ \hline
1 & 4/6 & 2/3 & 6/12 \\ \hline
2 & 0/6 & 0/3 & 3/12 \\ \hline
3 & 0/6 & 1/3 & 5/12 \\ \hline
Global & 4/6 & 3/3 & 14/12 \\ \hline
1-Global & 0 & -1/3 & -6/12 \\ \hline
\end{tabular}

\label{tab:resultsDescriptors}
\end{table}

Here a review of the results suggests that Deck and Card are both equally appropriate places to search for the bug reported in descriptor one. The results also suggest that the matches for  descriptor one, in Deck, contribute all of the matches that that constructor contributes to the global query. In contrast, the matches for descriptor one in Card contribute only two-thirds of the matches that that constructor contributes to the global query. Hence we can say that Card is slightly less specific to descriptor one than Deck, and so the ‘Sift’ component suggests that constructor one (Deck) is more appropriate for descriptor one than Card: as is appropriate for the shuffling bug.

\section{Empirical Design} \label{sec4:design}

\begin{figure*}[htbp]
\centerline{\includegraphics[width=0.89\textwidth]{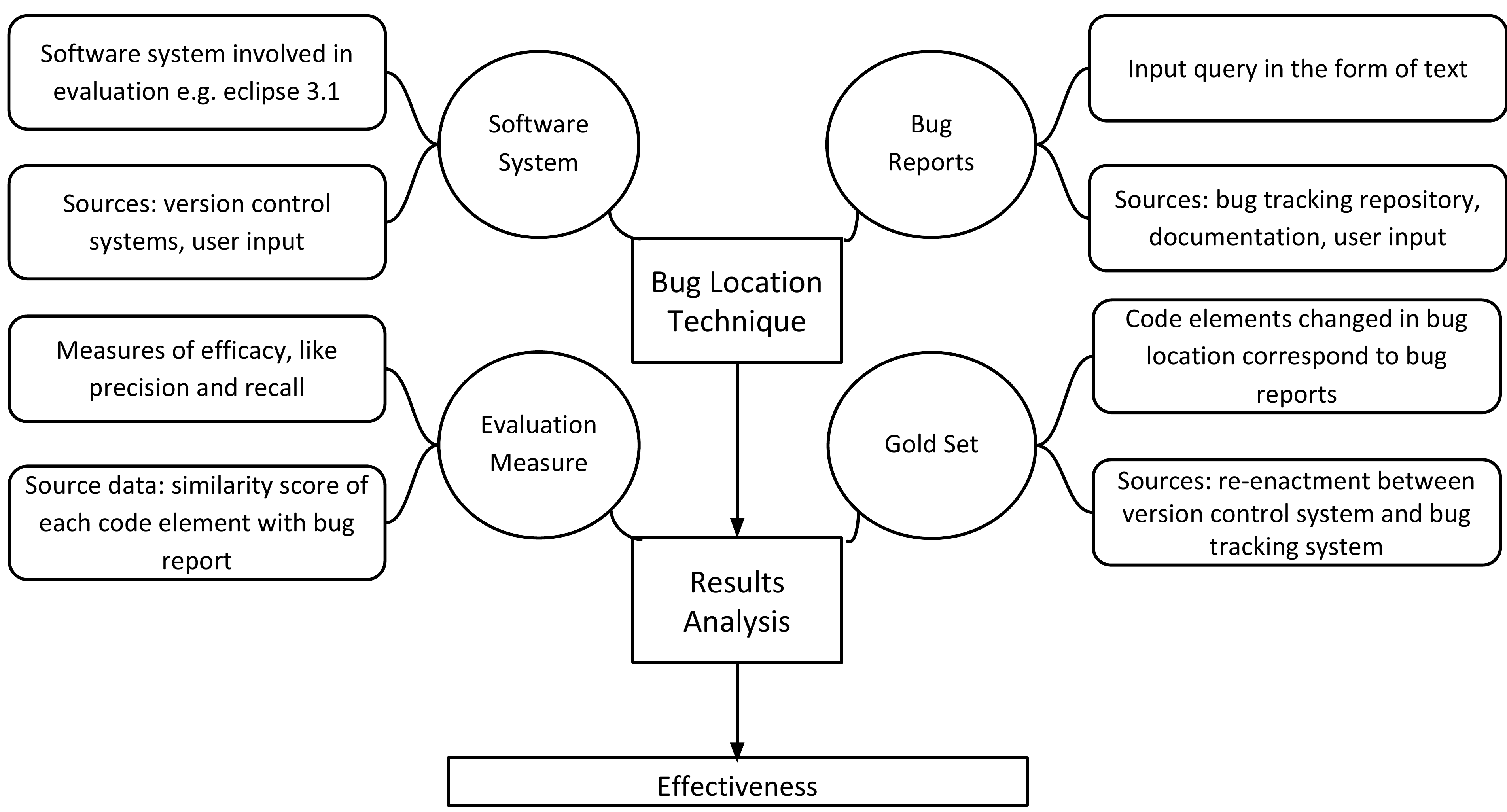}}
\caption{Empirical Components to Evaluate a Bug Localization Technique \cite{razzaq2019empirical}}
\label{fig1:Components}
\end{figure*}

In order to find the best-of-breed bug-localization techniques, empirical evaluations need to be performed. Figure \ref{fig1:Components} presents the empirical components that can be used to evaluate an IR-based bug localization technique \cite{razzaq2018state}. The essential inputs to an Information Retrieval (IR) based bug localization technique are the bug reports (or other textual queries) and a software system (under investigation). However, to characterize the empirical design of bug localization techniques these inputs can be expanded to include the evaluation measures employed to quantify the results of a technique and the actual source code elements changed in response to a set of bug reports. The latter is known as the gold set \cite{razzaq2018state}. The source code elements changed for a set of bug reports are often obtained from version control systems using a re-enactment process, where the bug reports already existing in bug tracking repositories are mapped to the locations of the code changed when the bug was addressed, as determined by the version control systems. 

While locating bugs, the localization techniques assign a similarity score to each source code element in a system for each bug. This means the output of bug localization techniques is a ranked list of source code elements for each bug report. In empirical evaluations, the ranked list is compared to the gold sets, in line with some evaluation measure, to quantify the quality of the results.

The empirical design employed in this paper is largely selected based on the existing state-of-the-art work in IR-based bug localization [5-7], and the details are now described:
\subsection{Comparison Techniques}
In this study, BoostNSift, is compared to openly-available, commonly-compared-to, IR-based bug localization techniques: BLIA, BLUiR, and BugLocator \cite{youm2017improved,zhou2012should,kim2013should,ye2016mapping,saha2013improving}. This selection is based on the \textit{baseline technique} criteria presented by \cite{razzaq2019empirical} et al. That is, in order to identify the best-of-breed, a new technique must be compared with openly-available, commonly-compared-to, fully-implemented and programmer-independent technique(s), where no intelligent assistance from the programmer is required.

\subsection{Dataset}

We use a dataset of 6851 bugs \cite{youm2017improved} from four popular open-source software systems to evaluate our technique against the other bug localization approaches. The four open-source projects are Eclipse.Platform.UI, AspectJ, SWT, and ZXing, and vary from small to large datasets, as illustrated in Table \ref{tab2:Systems}. Zhou's study \cite{zhou2012should} employed a dataset based on bugs from these four open-source systems (using Eclipse rather than Eclipse.Platform.UI), and the online repository makes that dataset available for three of the systems (not Eclipse). But the source code associated with the individual bugs is reported on at file (not method) level. Youm, in his evaluation of BLIA \cite{youm2017improved}, refined Zhou's dataset to method-level granularity, but again not for Eclipse.Platform.UI, and he only reported the method names: not the source code or identifiers associated with each method. We used the information in Youm's article to extract the relevant source code associated with each method to create the dataset used to evaluate BoostNSift. 
\begin{table}[]
\centering
\caption{Dataset Description (* Eclipse.Platform.Ui)}
\label{tab2:Systems}
\setlength{\tabcolsep}{2pt}
\renewcommand{\arraystretch}{1.3}
\begin{tabular}{|l|c|c|c|}
\hline
\rowcolor[HTML]{C0C0C0} 
            & \textbf{\# Bugs} & \textbf{Time-Period} & \textbf{\# Methods} \\ \hline \hline
Eclipse.UI* & 6480            & 10/2001 – 01/2014    & 63089              \\ \hline
AspectJ     & 272             & 12/2002 – 07/2007    & 33972              \\ \hline
SWT         & 83              & 04/2002 – 12/2005    & 10987              \\ \hline
ZXing       & 16              & 03/2010 – 09/2010    & 2079               \\ \hline
\end{tabular}
\end{table}

In terms of the Eclipse.Platform.UI dataset,Ye et al.\cite{ye2016mapping} made open a version of the dataset employed by Zhou \cite{youm2017improved}. It was expanded in that it encompassed bugs from 2001-2014. We created the analogous method-level dataset adopting the guidelines presented by Youm et al.\cite{youm2017improved} and Zhou et al.\cite{zhou2012should}. That is, we scanned through the log messages of bugs and removed "false positives" not existent in the bug tracking repository. Then, we established links between bug reports and commits using the commit-to-bug fine-grained mapping already provided by Ye et al.\cite{ye2016mapping}. Then, to identify the associated method(s), the lines-of-code-changed information obtained was associated with the relevant methods. Finally, we validated the accuracy and completeness of all the datasets using the status of the bug reports (open/close) and the availability of the gold sets for each bug report. While checking the validity, we excluded 12 bugs from AspectJ, 15 from SWT, and 4 from ZXing for which the method level gold sets were not present. Table \ref{tab2:Systems} presents the final bugs selected after validation. The complete dataset is made available to researchers through the following link: https://github.com/razi-rechan/boostNsift.

\subsection{Evaluation Measures} \label{sec4:eval-measures}

Typically, IR-based bug localization techniques return all of the code elements in a resultant ranked-list, hence the precision and recall are largely dependent on a ratio of total code elements to buggy elements \cite{razzaq2019empirical, chochlov2017historical}. Therefore, we use three size-independent evaluation measures namely Mean Reciprocal Rank (MRR), Mean Average Precision (MAP), and Top-N to evaluate the effectiveness of BoostNSift. These evaluation measures are defined as follows:
\subsubsection{Mean Average Precision (MAP)} Average Precision (AP) measures the extent to which a bug localization technique places correctly-retrieved buggy elements towards the top of the ranked-list by calculating the precision value at each position of the ranked-list and then averaging the values. MAP measures the mean of the average precisions calculated for a set of bug reports. It can be represented as follows:
\begin{equation}
    \frac{1}{|BR|}\sum_{br=1}^{BR}{\frac{\sum_{r=1}^{N}{(P(r) * isRelevant(r))}}{|RelevantElements_{br}|}}
\end{equation}

where \textit{BR} is a set of bug reports and \textit{r} is the rank position of a retrieved source code element in the ranked-list containing retrieved results of \textit{N} such elements given a bug report \textit{br}. \textit{isRelevant(r)} is a binary function that assigns 1 to the rank position \textit{r} if it contains a buggy code element, and 0 otherwise, and \textit{P(r)}  is the function that computes precision after truncating the list immediately below the ranked position \textit{r}.  In this way, MAP assigns a higher precision score to buggy elements at the top of the ranked-list (where the denominator is small) than to the buggy elements at the bottom (where the denominator is larger). 

\subsubsection{Mean Reciprocal Rank (MRR)} Reciprocal Rank (RR) measures the reciprocal of the rank position  of the first buggy element found in the ranked-list retrieved by a bug localization technique. MRR measures the mean of the reciprocal ranks given a set of bug reports. MRR is defined as follows:
\begin{equation}
    \frac{1}{|BR|}\sum_{br=1}^{BR}{\frac{1}{ranq_{br}}}
\end{equation}

Where $rank_{br}$ is the rank position of the top buggy element calculated against a bug report \textit{br} from a set of bug reports \textit{BR}. In this way, MRR assesses the techniques for best-case analysis in ranking one buggy element towards the top of the ranked-list, providing a foothold to the bug's location.

\subsubsection{Top-N} This evaluation measure calculates the number of bug reports for which at least one element was found and ranked in the top N positions of returned results. For example, given a bug report, if at least one method in which the bug is fixed is found in the top N results, we consider the bug to be localized successfully in the top N rank. This metric emphasizes early precision over the total ranked count. For Top-N we used four different versions: Top-1, Top-5, Top-10, and Top-20, in line with the design conventions commonly used in the field \cite{youm2017improved,zhou2012should,kim2013should,ye2016mapping,saha2013improving}.

\subsection{Research Questions} The following research questions are used to assess the performance of BoostNSift in this paper:

\begin{itemize}
    \item \emph{RQ1 -- How does BoostNSift perform compared to existing, openly-available state-of-the-art techniques like BugLocator, BLUiR, and BLIA (using MAP, MRR, Top-1, Top-5, Top-10, and Top-20 measures, on Eclipse.UI, AspectJ, SWT, and ZXing datasets)?}
    
\end{itemize}

The BLIA technique \cite{youm2017improved} integrates several types of analyses by utilizing stack traces, and comments in bug reports, the structured information of source files, and the source code change history. The BLUiR technique \cite{saha2013improving} embodies structured IR, based on code constructs and combines that with an inter-bug similarity score. Finally, BugLocator \cite{zhou2012should} used a refined version of the traditional Vector Space Model (VSM) for IR and combines that with inter-bug similarity and file length normalization scores.

An extended version of BLIA for method level granularity is made available by Youm et al.\cite{youm2017improved}. However, implementations of BugLocator and BLUiR at method level are not available. To compare those techniques at method level, we employed their original implementations but extended them to method level. Specifically, from their built term-to-file similarity matrix we retrieved the vector representation for the terms in methods and bug reports and, then performed the analysis on those vectors \cite{zhang2019finelocator}.

Thus, to answer RQ1, we compare BoostNSift with BugLocator, BLUiR, and BLIA using all three evaluation measures. In this comparison, we employed the best performing configuration of BLIA (BLIA requires the setting of several parameters such as the relative weighting and IR and inter-bug similarity), in line with Youm et al.\cite{youm2017improved}. For BugLocator and BLUiR, we used the recommended settings in their corresponding studies \cite{youm2017improved,zhou2012should,saha2013improving}. For example, for BLUiR, we used k=1.0 and b=0.3 for their BM25 IR model. For BoostNSift, we used the recommended IR settings of those parameters for large to medium-sized systems (i.e. Eclipse.UI, AspectJ, and SWT), but for the small system (ZXing) the terms are likely to saturate earlier. Hence we reduced k to zero.

In the case of BLIA, results of two different versions of the technique are compared: BLIA(10-Files) is where first the file-level analysis was performed. Then the 10 top files were selected, and then method-level analysis was performed on those files. Then the top 10\% of the methods were selected for evaluation, Youm et al.\cite{youm2017improved} in their paper presented the results of this version. BLIA(Full) is the version where after performing the file and method level analysis all the results were selected excepting those with zero scores. To compare BoostNSift with BLIA(10-Files), we also generate a version of BoostNSift named BoostNSift(10\%) where we selected the top 10\% methods from the overall results of our technique.

\begin{itemize}
    \item \emph{RQ2 -- Do both components (QueryBooster and CodeSifter) of BoostNSift contribute toward its overall performance?}
    
\end{itemize}

To answer RQ2, the effect of each component on the performance of BoostNSift is assessed separately. To do that, we simply drop one component (i.e., QueryBooster or CodeSifter) from BoostNSift one-at-a-time and evaluate their individual performance. In this process, we create two variants of BoostNSift: BNS\_QB, and BNS\_CS. BNS\_QB combines QueryBooster with BM25 and excludes the CodeSifter part whereas BNS\_CS combines CodeSifter with BM25 and excludes the QueryBooster.

\begin{itemize}
    \item \emph{RQ3 -- How does BoostNSift compared to the BugLocator, BLUiR, and BLIA in terms of time efficiency on Eclipse.UI, AspectJ, SWT, and ZXing?}
    
\end{itemize}

Our hypothesis here was that BoostNSift would outperform the others due to its single-analysis sifting. To answer RQ3, we compare the time in minutes BoostNSift, BugLocator, BLUiR, and BLIA techniques took when executed for Eclipse.UI, AspectJ, SWT, and ZXing systems. This analysis is performed using an E5570 Dell machine with 16 GB RAM and I7, 2.60 GHz processor. While performing this analysis, we made sure that only a single user-level process was running on the machine.

\section{Results} \label{sec5:results}
\subsection{Comparison of Bug Localization Techniques}

For RQ1, Table \ref{tab4:Results} shows the comparison results for the method-level IR-based bug localization techniques.  The best performing solution for each measure, for each system, is presented in bold where BoostNSift(10\%) is compared with BLIA(10-Files) and BoostNSift is compared with other all. Since the BLIA(10-Files) results for SWT, AspectJ and ZXing are taken from the original study of Youm et al.\cite{youm2017improved} where they did not perform the Top-20 analysis, the Top-20 results are omitted from the table. Note that other than BLIA(10-Files) all evaluation results are obtained re-running the experiments. The figures highlighted in bold clearly show that the both versions of BoostNSift outperform their corresponding versions of BLIA and indeed the other comparison techniques BLUiR and BugLocator. For example, in terms of MRR score, BoostNSift(10\%) outperforms BLIA(10-Files) by 70\%, 26\%, 15\%, and 4\% on SWT, AspectJ, Eclipse.UI, and ZXing, respectively. Similar improvement in results in terms of MAP and other measures can also be observed in the table. The only exception, in terms of BoostNSift outperforming other techniques, is the Top-5, Top-10, and Top-20 results for ZXing, where BLIA or BugLocator outperformed BoostNSift. 

\begin{table}[]
\centering
\caption{Comparison Results of BoostNSift with BLIA, BLUiR, and BugLocator}
\label{tab4:Results}
\setlength{\tabcolsep}{2pt}
\renewcommand{\arraystretch}{1.3}
\begin{tabular}{llcccccc}
\hline
\rowcolor[HTML]{D9D9D9} 
\multicolumn{1}{|c|}{\cellcolor[HTML]{D9D9D9}\textbf{Systems}} & \multicolumn{1}{c}{\cellcolor[HTML]{D9D9D9}\textbf{Techniques}} & \textbf{MRR} & \textbf{MAP} & \textbf{Top1} & \textbf{Top5} & \textbf{Top10} & \textbf{Top20} \\ \hline \hline
\cellcolor[HTML]{F2F2F2} & \textbf{BoostNSift(10\%)} & \textbf{0.289} & \textbf{0.203} & \textbf{18} & \textbf{28} & \textbf{37} & \textbf{45} \\ \cline{2-8} 
\cellcolor[HTML]{F2F2F2} & \textbf{BLIA(10-Files)} & 0.170 & 0.100 & 10 & 20 & 32 & \begin{tabular}[c]{@{}c@{}}Not\\ published\end{tabular} \\ \cline{2-8} 
\cellcolor[HTML]{F2F2F2} & \textbf{BoostNSift} & \textbf{0.290} & \textbf{0.183} & \textbf{18} & \textbf{28} & \textbf{37} & \textbf{45} \\ \cline{2-8} 
\cellcolor[HTML]{F2F2F2} & \textbf{BLIA(Full)} & 0.132 & 0.070 & 5 & 22 & 32 & 41 \\ \cline{2-8} 
\cellcolor[HTML]{F2F2F2} & \textbf{BugLocator} & 0.073 & 0.044 & 3 & 10 & 16 & 23 \\ \cline{2-8} 
\multirow{-6}{*}{\cellcolor[HTML]{F2F2F2}\textbf{SWT}} & \textbf{BLUiR} & 0.072 & 0.041 & 1 & 13 & 20 & 24 \\ \hline \hline
\cellcolor[HTML]{F2F2F2} & \textbf{BoostNSift(10\%)} & \textbf{0.156} & \textbf{0.101} & \textbf{626} & \textbf{1394} & \textbf{1776} & \textbf{2193} \\ \cline{2-8} 
\cellcolor[HTML]{F2F2F2} & \textbf{BLIA(10-Files)} & 0.136 & 0.096 & 601 & 1294 & 1594 & 1976 \\ \cline{2-8} 
\cellcolor[HTML]{F2F2F2} & \textbf{BoostNSift} & \textbf{0.156} & \textbf{0.080} & \textbf{626} & \textbf{1394} & \textbf{1776} & \textbf{2194} \\ \cline{2-8} 
\cellcolor[HTML]{F2F2F2} & \textbf{BLIA(Full)} & 0.114 & 0.066 & 398 & 673 & 981 & 1233 \\ \cline{2-8} 
\cellcolor[HTML]{F2F2F2} & \textbf{BugLocator} & 0.009 & 0.006 & 21 & 82 & 126 & 179 \\ \cline{2-8} 
\multirow{-6}{*}{\cellcolor[HTML]{F2F2F2}\textbf{Eclipse.UI}} & \textbf{BLUiR} & 0.01 & 0.006 & 24 & 98 & 139 & 194 \\ \hline \hline
\cellcolor[HTML]{F2F2F2} & \textbf{BoostNSift(10\%)} & \textbf{0.139} & \textbf{0.100} & \textbf{24} & \textbf{49} & \textbf{63} & \textbf{73} \\ \cline{2-8} 
\cellcolor[HTML]{F2F2F2} & \textbf{BLIA(10-Files)} & 0.110 & 0.080 & 17 & 43 & 59 & \begin{tabular}[c]{@{}c@{}}Not\\ published\end{tabular} \\ \cline{2-8} 
\cellcolor[HTML]{F2F2F2} & \textbf{BoostNSift} & \textbf{0.140} & \textbf{0.089} & \textbf{24} & \textbf{49} & \textbf{63} & \textbf{73} \\ \cline{2-8} 
\cellcolor[HTML]{F2F2F2} & \textbf{BLIA(Full)} & 0.138 & 0.063 & 23 & 44 & 60 & 71 \\ \cline{2-8} 
\cellcolor[HTML]{F2F2F2} & \textbf{BugLocator} & 0.058 & 0.035 & 6 & 22 & 29 & 46 \\ \cline{2-8} 
\multirow{-6}{*}{\cellcolor[HTML]{F2F2F2}\textbf{AspectJ}} & \textbf{BLUiR} & 0.033 & 0.016 & 1 & 10 & 28 & 53 \\ \hline \hline
\cellcolor[HTML]{F2F2F2} & \textbf{BoostNSift(10\%)} & \textbf{0.208} & \textbf{0.208} & 3 & 4 & 4 & 5 \\ \cline{2-8} 
\cellcolor[HTML]{F2F2F2} & \textbf{BLIA(10-Files)} & 0.200 & 0.200 & 3 & \textbf{5} & \textbf{6} & \begin{tabular}[c]{@{}c@{}}Not\\ published\end{tabular} \\ \cline{2-8} 
\cellcolor[HTML]{F2F2F2} & \textbf{BoostNSift} & \textbf{0.213} & \textbf{0.172} & 3 & 4 & 4 & 6 \\ \cline{2-8} 
\cellcolor[HTML]{F2F2F2} & \textbf{BLIA(Full)} & 0.196 & 0.159 & 3 & \textbf{5} & \textbf{6} & 6 \\ \cline{2-8} 
\cellcolor[HTML]{F2F2F2} & \textbf{BugLocator} & 0.117 & 0.093 & 1 & 4 & 4 & \textbf{7} \\ \cline{2-8} 
\multirow{-6}{*}{\cellcolor[HTML]{F2F2F2}\textbf{ZXing}} & \textbf{BLUiR} & 0.113 & 0.096 & 1 & 3 & 4 & 5 \\ \hline
\end{tabular}
\end{table}

Table \ref{tab4:Statistics} reports the results of the Wilcoxon Rank-Sum test (p-value) and Cliff's delta. The results show that the performance of BoostNSift is statistically significant at an $\alpha$ of 0.001 for SWT, AspectJ, and Eclipse.UI for both MRR and MAP. However, in the case of ZXing, the performance of BoostNSift is not statistically different from BLIA. This may be because the condition of minimum data-points recommended for analysis \cite{green1991many} is not met in the case of ZXing but, if anything, on this smaller system the BLIA alternatives seem to outperform BoostNSift. (The results of the statistical analysis are not presented for Top-N results because such an analysis is not possible on one datum. However, the table suggests that there is no trend regardless).
\begin{table}[]
\centering
\caption{Statistical analysis results of the Wilcoxon Rank-Sum test and Cliff's delta}
\label{tab4:Statistics}
\setlength{\tabcolsep}{2pt}
\renewcommand{\arraystretch}{1.3}
\begin{tabular}{llllcc}
\hline
\rowcolor[HTML]{D9D9D9} 
\multicolumn{1}{c}{\cellcolor[HTML]{D9D9D9}\textbf{Systems}} & \multicolumn{1}{c}{\cellcolor[HTML]{D9D9D9}\textbf{Measures}} & \multicolumn{1}{c}{\cellcolor[HTML]{D9D9D9}\textbf{Techniques}} & \multicolumn{1}{c}{\cellcolor[HTML]{D9D9D9}\textbf{P-Value}} & \textbf{d} & \textbf{Effect Size} \\ \hline \hline
\cellcolor[HTML]{F2F2F2} &  & \textbf{vs. BLIA} & \textbf{0.000} & 0.485 & large \\ \cline{3-6} 
\cellcolor[HTML]{F2F2F2} &  & \textbf{vs. BLUiR} & \textbf{0.000} & 0.505 & large \\ \cline{3-6} 
\cellcolor[HTML]{F2F2F2} & \multirow{-3}{*}{\textbf{MRR}} & \textbf{vs. BugLocator} & \textbf{0.000} & 0.495 & large \\ \cline{2-6} 
\cellcolor[HTML]{F2F2F2} &  & \textbf{vs. BLIA} & \textbf{0.000} & 0.542 & large \\ \cline{3-6} 
\cellcolor[HTML]{F2F2F2} &  & \textbf{vs. BLUiR} & \textbf{0.000} & 0.587 & large \\ \cline{3-6} 
\multirow{-6}{*}{\cellcolor[HTML]{F2F2F2}\textbf{SWT}} & \multirow{-3}{*}{\textbf{MAP}} & \textbf{vs. BugLocator} & \textbf{0.000} & 0.575 & large \\ \hline \hline
\cellcolor[HTML]{F2F2F2} &  & \textbf{vs. BLIA} & \textbf{0.000} & 0.348 & medium \\ \cline{3-6} 
\cellcolor[HTML]{F2F2F2} &  & \textbf{vs. BLUiR} & \textbf{0.000} & 0.505 & large \\ \cline{3-6} 
\cellcolor[HTML]{F2F2F2} & \multirow{-3}{*}{\textbf{MRR}} & \textbf{vs. BugLocator} & \textbf{0.000} & 0.487 & large \\ \cline{2-6} 
\cellcolor[HTML]{F2F2F2} &  & \textbf{vs. BLIA} & \textbf{0.000} & 0.403 & medium \\ \cline{3-6} 
\cellcolor[HTML]{F2F2F2} &  & \textbf{vs. BLUiR} & \textbf{0.000} & 0.513 & large \\ \cline{3-6} 
\multirow{-6}{*}{\cellcolor[HTML]{F2F2F2}\textbf{Eclipse.UI}} & \multirow{-3}{*}{\textbf{MAP}} & \textbf{vs. BugLocator} & \textbf{0.000} & 0.511 & large \\ \hline \hline
\cellcolor[HTML]{F2F2F2} &  & \textbf{vs. BLIA} & \textbf{0.000} & 0.334 & medium \\ \cline{3-6} 
\cellcolor[HTML]{F2F2F2} &  & \textbf{vs. BLUiR} & \textbf{0.000} & 0.481 & large \\ \cline{3-6} 
\cellcolor[HTML]{F2F2F2} & \multirow{-3}{*}{\textbf{MRR}} & \textbf{vs. BugLocator} & \textbf{0.000} & 0.347 & medium \\ \cline{2-6} 
\cellcolor[HTML]{F2F2F2} &  & \textbf{vs. BLIA} & \textbf{0.000} & 0.23 & small \\ \cline{3-6} 
\cellcolor[HTML]{F2F2F2} &  & \textbf{vs. BLUiR} & \textbf{0.000} & 0.487 & large \\ \cline{3-6} 
\multirow{-6}{*}{\cellcolor[HTML]{F2F2F2}\textbf{AspectJ}} & \multirow{-3}{*}{\textbf{MAP}} & \textbf{vs. BugLocator} & \textbf{0.000} & 0.26 & small \\ \hline \hline
\cellcolor[HTML]{F2F2F2} &  & \textbf{vs. BLIA} & 0.388 & 0.082 & negligible \\ \cline{3-6} 
\cellcolor[HTML]{F2F2F2} &  & \textbf{vs. BLUiR} & \textbf{0.015} & 0.161 & small \\ \cline{3-6} 
\cellcolor[HTML]{F2F2F2} & \multirow{-3}{*}{\textbf{MRR}} & \textbf{vs. BugLocator} & \textbf{0.037} & 0.155 & small \\ \cline{2-6} 
\cellcolor[HTML]{F2F2F2} &  & \textbf{vs. BLIA} & 0.388 & 0.089 & negligible \\ \cline{3-6} 
\cellcolor[HTML]{F2F2F2} &  & \textbf{vs. BLUiR} & \textbf{0.043} & 0.151 & small \\ \cline{3-6} 
\multirow{-6}{*}{\cellcolor[HTML]{F2F2F2}\textbf{ZXing}} & \multirow{-3}{*}{\textbf{MAP}} & \textbf{vs. BugLocator} & \textbf{0.047} & 0.147 & small \\ \hline 
\end{tabular}%
\end{table}
The effect sizes (d) are large or medium in most of the comparisons using SWT, AspectJ and Eclipse.UI. The effect sizes of MAP and MRR between BoostNSift and comparator techniques are particularly all large in the case of SWT. However, the picture is less clear in the case of ZXing. There may be a size issue with ZXing (and ZXing's gold-set) being considerably smaller than the other systems \cite{razzaq2021effect}. To evaluate this, we will assess BoostNSift against other small systems in the future but it should be noted that, intuitively, the problem of bug localization is exacerbated in larger systems, where BoostNSift seems to perform better. 

\textbf{\emph{Answer to RQ1 --}} From the techniques' comparison results presented in Table \ref{tab4:Results} and analysis results presented in Table \ref{tab4:Statistics}, we can conclude that BoostNSift performs significantly better than different versions of BLIA, BugLocator, and BLUiR using MAP, MRR evaluation measures for the larger systems studied: SWT, AspectJ, and Eclipse.UI. However, the results for the smaller ZXing system, particularly with respect to Top-N, are not clear-cut. 
\subsection{Effect of Individual Components}
To demonstrate the effect of individual components, Table \ref{tab6:Components} shows the results of two different versions of BoostNSift i.e. BNS\_QB and BNS\_CS. The results exhibit the importance of each component. The results suggest that the BNS\_QB version largely preserves the performance in terms of MRR. In fact, it matches the overall BoostNSift performance on SWT and improves it on AspectJ and ZXing. BNS\_CS, on the other hand, decreases the MRR performance in each instance but does do better than BNS\_QB in terms of getting nearer the overall BoostNSift technique in terms of MAP on larger systems: that is, in terms of an improved buggy to non-buggy element ratio for each position of the ranked-list. 
\begin{table}[]
\centering
\caption{Effect of each Component on the Performance of BoostNSift}
\label{tab6:Components}
\renewcommand{\arraystretch}{1.3}
\begin{tabular}{llcc}
\hline 
\rowcolor[HTML]{D9D9D9} 
\multicolumn{1}{c}{\cellcolor[HTML]{D9D9D9}\textbf{Systems}} & \multicolumn{1}{c}{\cellcolor[HTML]{D9D9D9}\textbf{Techniques}} & \textbf{MRR} & \textbf{MAP} \\ \hline \hline
\cellcolor[HTML]{F2F2F2} & \textbf{BoostNSift} & 0.29 & 0.183 \\ \cline{2-4} 
\cellcolor[HTML]{F2F2F2} & \textbf{BNS\_QB} & 0.29 & 0.146 \\ \cline{2-4} 
\multirow{-3}{*}{\cellcolor[HTML]{F2F2F2}\textbf{SWT}} & \textbf{BNS\_CS} & 0.27 & 0.171 \\ \hline \hline
\cellcolor[HTML]{F2F2F2} & \textbf{BoostNSift} & 0.156 & 0.101 \\ \cline{2-4} 
\cellcolor[HTML]{F2F2F2} & \textbf{BNS\_QB} & 0.138 & 0.057 \\ \cline{2-4} 
\multirow{-3}{*}{\cellcolor[HTML]{F2F2F2}\textbf{Eclipse.UI}} & \textbf{BNS\_CS} & 0.126 & 0.062 \\ \hline \hline
\cellcolor[HTML]{F2F2F2} & \textbf{BoostNSift} & 0.14 & 0.089 \\ \cline{2-4} 
\cellcolor[HTML]{F2F2F2} & \textbf{BNS\_QB} & 0.142 & 0.063 \\ \cline{2-4} 
\multirow{-3}{*}{\cellcolor[HTML]{F2F2F2}\textbf{AspectJ}} & \textbf{BNS\_CS} & 0.137 & 0.084 \\ \hline \hline
\cellcolor[HTML]{F2F2F2} & \textbf{BoostNSift} & 0.208 & 0.208 \\ \cline{2-4} 
\cellcolor[HTML]{F2F2F2} & \textbf{BNS\_QB} & 0.237 & 0.136 \\ \cline{2-4} 
\multirow{-3}{*}{\cellcolor[HTML]{F2F2F2}\textbf{ZXing}} & \textbf{BNS\_CS} & 0.218 & 0.115 \\ \hline
\end{tabular}
\end{table}

\textbf{\emph{Answer to RQ2 --}} Table \ref{tab6:Components} suggests that QueryBooster contributes mostly to the performance of BoostNSift in terms of MRR but at the expense of MAP. CodeSifter is less effective in terms of MRR but does retain more of the overall technique's MAP performance. In summary, the overall technique seems a stronger technique when both MAP and MRR are important measures for the user, but the QueryBooster variant should possibly be considered in isolation when only MRR is sought.

\subsection{Simplicity of BoostNSift}

Table \ref{tab7:Simplicity} compares the time each technique took when executed on the four systems. The table shows that BoostNSift is more efficient than BugLocator, BLUiR, and BLIA in that it took less time. For example, BLIA(Full) took 852 minutes more (i.e. 539\% higher) than BoostNSift when executed on Eclipse.UI.
\begin{table}[]
\centering
\caption{Time in Minutes Taken by each Technique}
\label{tab7:Simplicity}
\begin{tabular}{lcccc}
\hline
\rowcolor[HTML]{D9D9D9} 
\multicolumn{1}{c}{\cellcolor[HTML]{D9D9D9}\textbf{}} & \textbf{Eclipse.UI} & \textbf{AspectJ} & \textbf{SWT} & \textbf{ZXing} \\ \hline \hline
\textbf{BoostNSift} & 158 & 15 & 7 & 0.25 \\ \hline
\textbf{BugLocator} & 480 & 110 & 18 & 4 \\ \hline
\textbf{BLUiR} & 492 & 111 & 18 & 4 \\ \hline
\textbf{BLIA(Full)} & 1010 & 180 & 21 & 5 \\ \hline
\end{tabular}
\end{table}

\textbf{\emph{Answer to RQ3 --}}  Table \ref{tab7:Simplicity} shows that BoostNSift is more efficient than BugLocator, BLUiR, and BLIA in that it took less time to return results from all four systems. Hence, in answer to RQ3, we can say that BoostNSift seems more time-efficient than the compared-to techniques.

\section{Discussion} \label{sec6:discussion}

\subsection{Effectiveness Aspects of BoostNSift}

Overall, Table \ref{tab4:Results} shows a significant improvement of BoostNSift results as compared to the BLIA \cite{youm2017improved}, BugLocator \cite{zhou2012should}, and BLUiR \cite{saha2013improving} on larger systems. For example, for the SWT system, the performance of BoostNSift(10\%) was 103\% and 70\% better than BLIA(10-Files) in terms of MAP and MRR, respectively. Similarly, BoostNSift outperformed BugLocator by 316\% and 297\% and BLUiR by 346\% and 303\% in terms of MAP and MRR, respectively, on the same system. These figures illustrate the effectiveness of BoostNSift over the three larger systems at method level and suggests real potential in adopting both pre-weighting-of-the-query and post-normalization-of-the-code steps in combination.

The exception is ZXing, where the MRR and MAP differences are small across techniques and, in the case of BLIA, there is no statistical difference observible. Indeed, for Top-N results, BLIA (both variants) and BugLocator performed better than BoostNSift for various `N's. This may be because the ZXing system is comparatively very small (only 2079 methods) and the evaluation carried out with very few bugs (16 in total). It is also possible that some system-level characteristics \cite{razzaq2019empirical} may have had an impact on the performance of BoostNSift. 

Another interpretation though, is that for small systems, a composite analysis (which includes file and method level) maybe more effective than singular analysis. It should also be pointed out however, that techniques that contribute to bug-location are most valuable in larger systems and the current results suggest that BoostNSift is most comfortable in that arena.

\subsection{Simplicity of BoostNSift}

Table \ref{tab7:Simplicity} shows that BoostNSift is more efficient than the compared-to state-of-the-art techniques in terms of time taken to locate bugs, a valuable attribute in its application to larger systems. This is because BoostNSift is a purely IR-based technique, which does not need any extra orthogonal analysis. This makes BoostNSift simpler than the recent proposed, more sophisticated techniques which typically need some additional input(s) from the history of a software system \cite{nichols2010augmented,youm2017improved,zhou2012should,zhang2016baha,zhang2019finelocator,wen2016locus,wang2014version} and correspondingly different analysis type(s) \cite{wang2014version,le2015information,shi2014empirical}. Given this relative simplicity of BoostNSift, we believe that there is still, some room for effectiveness improvement in BoostNSift, in terms of integrating it into hybrid techniques.

\subsection{Contribution of Individual Components}

When comparing BoostNSift to its individual components, the contribution of both components is obvious: without the query booster, both MRR and MAP results fall off significantly and without the code sifter the MAP results fall off significantly. One might then address the scenario where just MRR results are important. But our results reveal no significant difference between BNS\_CS and the more holistic BoostNSift in terms of MRR. So, a sensible strategy seems to be to apply BoostNSift holistically on larger software systems.   

\section{Threats to Validity} \label{sec7:threats}

This section presents several threats that may potentially impact the validity of our study.

\subsection{Construct Validity}
A core construct validity issue in bug location techniques evaluation is the quality of the benchmark against which the techniques are compared. In this paper, the selected benchmark is created using re-enactment where bug reports in issue-tracking repositories are mapped to the changed methods in version control systems.  However, some of the links between bug-reports and their fixing commits may be less accurate because those bugs can reopen in the future or code can be altered due to other subsequent changes. To address this issue partially, we gathered datasets that have been validated by several researchers and employed by several studies \cite{zhou2012should,saha2013improving,youm2017improved} But still, there is no guarantee of absolute completeness and accuracy of the dataset. 

Another concern that can threaten construct validity is the evaluation measures used. The employed measures are commonly used for the evaluation of IR-based bug localization techniques \cite{youm2017improved,zhang2019finelocator,tantithamthavorn2018impact,rahman2016improved}. However, they capture the effectiveness perspective of evaluation and not necessarily other important attributes like usability \cite{razzaq2019empirical}.
\subsection{Internal Validity}
In this paper, internal validity threats are related to the implementation of BoostNSift and pre-processing of its inputs. To configure the queries for boosting, BoostNSift relied on the state-of-the-art and was tested for several configurations based on the different weights of the title, description, or comments of a bug report, for one system. For each part of the bug, the range of weights tested varied from 0.5 to 4.0 with step-size 0.1. To keep consistent settings, we adopted the weighting scheme that performed well on AspectJ system and was reported in literature (see Table \ref{tab3:Setting}).  For the pre-processing of the code elements and queries, we also employed best-identified practice identified in several studies \cite{dit2015configuring,biggers2014configuring,tantithamthavorn2018impact,moreno2015query,thomas2013impact} (see Section~\ref{sec3.1:qb}). However, a better configuration may exist and this could have negatively impacted on the BoostNSift results reported. Indeed, if a specific software system has a substantial repository of bug reports and associated commits, the possibility of identifying a system-specific optimal weighting would be possible, further improving the performance of BoostNSift. 

As regards the implementations of techniques, for BM25, we used the existing implementation of Lucene which is extensively used by different tools \footnote{https://lucene.apache.org/solr/} and techniques \cite{thomas2013impact,razzaq2019empirical,ali2013trustrace}. Similarly, we adapted the existing implementation of BLIA and BugLocator \cite{zhou2012should,youm2017improved}. However, changing the granularity to method level for these two latter techniques may have caused small differences in those techniques' performance. Finally, excepting the ZXing system, we employed the recommended IR settings of parameters for BM25. However, better settings may exist particular for the bug localization for a given system. 
\subsection{External Validity}
The paper's use of existing state-of-the-art data-sets may have caused some external validity threats to our results, in that we only used four open-source Java-based systems. In the future, we plan to reduce this threat to external validity by investigating additional software systems, including those written in programming languages like C, C++ and Python.

\section{Conclusions and Future Work} \label{sec8:conc}
Software bugs are pervasive in software development which makes bug localization a frequent, effort-intensive and resource-consuming activity. This paper proposes a novel bug localization technique called BoostNSift that performs at method-level granularity. It embeds query boosting and code sifting in conjunction with the BM25 Information Retrieval (IR) model. In the query boosting step, it adds weight to the title field. In code sifting, BoostNSift compares the relevance of a code element for a specific bug report to its relevance for an aggregation of bug reports. For a specific bug, the aim of this latter step is to filter out the code elements less likely to be specifically related to the searched-for bug from the returned, ranked list without learning to rank \cite{ye2016mapping,ali2013trustrace} or relying on any additional inputs \cite{poshyvanyk2007feature,dit2013integrating}. Experiments on four open-source systems demonstrate that the proposed BoostNSift can produce better performance than existing, commonly-compared bug localization techniques (BugLocator, BLUiR, and BLIA) in terms of MRR and MAP. 

In the future, we plan to extend this work by exploiting the method structure (method names, variables, comments) in our `boost' phase. Also, we want to extend BoostNSift with a machine learning technique that automatically identifies the best boosting configurations for each part of the query based on historical, system-specific information. 

In this paper, we limit the evaluation software systems to commonly used (for evaluation) Java-based systems. However, bug-location techniques should not to be limited to a particular programming language or OS systems. Hence, in the future, we want to test BoostNSift for C/C++ based and commercial systems. Finally, since BoostNSift currently only uses bug reports and systems as inputs, we believe that room still exists towards its further improvement using complimentary techniques e.g. by using stack trace information \cite{le2015information}, bug similarity information \cite{zhou2012should} and other historical information \cite{youm2017improved,wang2014version}. 
\section*{Acknowledgment}

Research supported, in part, by Science Foundation Ireland grant 13/RC/2094. 

\vspace{12pt}

\bibliographystyle{plain}      
\bibliography{ICPCbib.bib}   

\end{document}